# New strategy for black phosphorus crystal growth through ternary clathrate

Sheng Li[1,*], Xiaoyuan Liu[1,*], Xing Fan[2,3], Yizhou Ni[4], John Miracle[5], Nikoleta Theodoropoulou[5], Jie Sun[3,6], Shuo Chen[4], Bing Lv[1,§], Qingkai Yu[2,§]

[1] Department of Physics, The University of Texas at Dallas, Richardson, Texas 75080
[2] Ingram School of Engineering, Texas State University, San Marcos, TX 78666
[3] Key Laboratory of Optoelectronics Technology, College of Electronic Information and Control Engineering, Beijing University of Technology, Beijing 100124, China
[4] TcSUH and Department of Physics, University of Houston, Houston, Texas 77204
[5] Department of Physics, Texas State University, San Marcos, TX 78666
[6] Quantum Device Physics Laboratory, Department of Microtechnology and Nanoscience, Chalmers University of Technology, Göteborg 41296, Sweden

* The first two authors contributed equally to this work.

§ To whom correspondence should be addressed: B. Lv: blv@utdallas.edu, and Q. Yu: qingkai.yu@txstate.edu.

**Abstract:** we are reporting a new synthetic strategy to grow large size black phosphorus (Black-P) crystals through a ternary clathrate $Sn_{24}P_{22-x}I_8$ under lower synthetic temperature and pressure. The Black-P crystals are found grown *in situ* at the site where the solid clathrate originally resides, which suggests chemical vapor mineralizer does not play a critical role for the Black-P formation. More detailed systematical studies has indicated the P vacancies in the framework of ternary clathrate $Sn_{24}P_{22-x}I_8$ is important for the subsequent Black-P from phosphorus vapors, and a likely Vapor-Solid-Solid (VSS) model is responsible for the Black-P crystal growth. The obtained room temperature mobility μ is ~350 cm$^2$/V·s from Hall measurements at mechanically-cleaved flake, where noticeable micro-cracks are visible. The obtained high mobility value further suggest the high quality of the Black-P crystals synthesized through this route.

**Introduction:**

Black phosphorus (Black-P) and its monolayer counterpart, phosphorene, emerge as a promising candidate of two-dimensional (2D) materials owing to their comprehensive outstanding properties, especially the range of its bandgaps and potentially high carrier mobilities.[1-4] Therefore, they attract tremendous immediate interests from condensed matter physicists, chemists, semiconductor device engineers, and material scientists. Black-P was firstly discovered more than one hundred year ago[5] and is the most stable allotrope of phosphorus (P) with puckered layer structure. The monolayer Black-P includes two atomic layers with two different kinds of P–P bonding patterns. The top view of Black-P along the z direction shows a hexagonal structure with bond angles of 96.3° and 102.1°, respectively. Recent theoretical studies have predicted that monolayer Black-P (phosphorene) can have an extremely high hole mobility (10,000 cm$^2$V$^{-1}$s$^{-1}$).[1] Depending on the thickness, the bandgap of Black-P can cover 0.3-2.0 eV,[2] which bridges the zero bandgap of graphene and the relatively large band gap of many transition metal dichalcogenides (TMDs). The bandgap of Black-P covers near and mid- infrared spectrum, thus making it an appealing candidate for near and mid- infrared optoelectronics.[6,7] Moreover, the mobility/on–off ratio combination for Black-P falls into a region that could not be easily covered by graphene or TMDs.

Black-P was historically synthesized under high pressure (~ 1.2 GPa) conditions from white phosphorus (White-P) and red phosphorus (Red-P).[5] It was also reported that Black-P was prepared from solution of White-P in mercury[8] or bismuth-flux.[8-10] Since readily stable Red-P does not dissolve in the flux, and the lack of high purity White-P due to its chemical activity, some sophisticated chemical treatment and apparatus are normally adopted for the synthesis, yet still yield deformed crystals, not to mention the potential high cost and high toxicity associated with the flux.[10-11] It was not until very recently that Black-P was reported to be synthesized at a lower pressure and much cleaner controlled atmosphere through mineralizer-assisted vapor transport reaction.[12] The mineralization additive reported initially was Au/Sn/SnI$_4$,[12] and then simplified to Sn/SnI$_4$.[13] Because the reaction temperature (~600 °C) is much higher than the sublimation temperature of SnI$_4$ (~ 200°C), it is further simplified to use Sn/I$_2$[14] to grow Black-P ribbons with a muffle furnace (with less temperature gradient and large reaction space). Zhao et al[14] further extended their studies to investigate the growth mechanism, and proposed some speculated growth models based on their results.

However, many questions regarding to Black-P crystal growth remain. For examples, what is the exact role of these mineralization additive to promote the formation of Black-P and their specific mechanism at different temperature stages? What exactly is the formation temperature range and growth model for Black-P? Is it a slow growing process from the highest temperature (>600 °C) down to the lowest temperature (~200 °C), similar to the crystallization and crystal growth in the flux? Or it is a rather rapid process that occurs at a much narrower temperature range? Different species such as Sn, I$_2$, SnI$_2$, SnI$_4$, and Au have been used in the past as necessary chemicals assisting the growth of Black-P. However, the exact functions of these chemicals are not well understood yet. At the reported highest temperature (650 °C or 590 °C), the most thermodynamically stable species is the SnI$_2$. Therefore, those additives such as Sn, I$_2$ or SnI$_4$ will probably either react with each other or decompose to form SnI$_2$(g): $Sn + I_2 \xrightarrow{>100°C} SnI_2$ ; $SnI_4 \xrightarrow{>340°C} SnI_2 + I_2$. The gaseous SnI$_2$ will react with P$_4$ vapor (sublimed from Red-P) upon cooling, which assists the formation of Black-P. Previous studies[14] have ambiguously speculated that some gaseous ternary P-Sn-I compound formed before the formation of Black-P ribbons is crucial for the nucleation and continuous growth of the Black-P crystals. However, the exact composition of the solid state materials prior to the formation of Black-P, and its role to promote the reaction, remain unknown. Under this motivation, we have carried out systematic studies to investigate the exact growth model of Black-P. We found out that the ternary type-I inverse clathrate Sn$_{24}$P$_{22-x}$I$_8$ is the sole ternary phase of Sn, P, and I. More importantly, we have grown large size an high quality Black-P crystals *in-situ* through this solid ternary inverse clathrate Sn$_{24}$P$_{22-x}$I$_8$ in a controlled manner, and clarified that ternary inverse clathrate Sn$_{24}$P$_{22-x}$I$_8$, not the commonly accepted vapor mineralizer, is critical for Black-P growth. We expect that the findings in this work could inspire community to design the new growth strategies for Black-P, and eventually pave a practical way for future large and high quality phosphorene film growth on different substrates.

## 2. Experimental Section:

**2.1 Synthesis of ternary inverse clathrate Sn$_{24}$P$_{22-x}$I$_8$:** A stoichiometric mixture of red phosphorus lump (Alfa Aesar, 99.999%), tin shot (Alfa Aesar, 99.9999%), and iodine crystals (Alfa Aesar, 99.9%) was sealed under vacuum in a silica tube. The tube was slowly (0.5 °C/min) heated to 550 °C, annealed for 5 days, and then furnace-cooled to room temperature by turning of the power. After regrinding, the sample was heated at the same conditions for another 5 days. The Sn$_{24}$P$_{22-x}$I$_8$ (x~2.7) is found to be stable and remains as solid below 520-550 °C, and decomposes irreversibly above 600 °C. Previous single crystal studies have shown[15-16] that the P vacancy in this material could be as high as 13% (x ~

2.7), which is also consistent with our scanning electron microscopy (SEM) energy-dispersive X-ray spectroscopy (EDX) analysis.

**2.2 Synthesis of Black-P crystals through ternary clathrate**: ~20 mg of $Sn_{24}P_{22-x}I_8$ powder are sealed in the silica tube under vacuum with 300 mg Red-P are loaded on top of the $Sn_{24}P_{22-x}I_8$ powder. The assembly was then put into the box furnace where the reactant end was at the center of the furnace, slowly heated up (0.5°C/min) to 550 °C for 20 hours where $Sn_{24}P_{22-x}I_8$ remained as solid, and then rapidly cooled down (in 10-20 hours) to 350 °C followed by furnace quenching of the sample. This time, layered square size of Black-P crystals formed at the exact place of the original ternary clathrate.

**2.3 Synthesis of Black-P through elements**: Red-P lump (Alfa Aesar, 99.999%), tin shot (Alfa Aesar, 99.9999%), and iodine crystals (Alfa Aesar, 99.9%) were used. Different Sn:I ratio and different Red-P amount were tested to further investigate the growth mechanism of Black-P. For a typical successful growth, the mixture of Red P:Sn:$I_2$ was sealed in silica ampoules, and was slowly heated up (0.5 °C/min) to target temperature (600-620 °C) in a muffle furnace/tube. The assembly was kept at this temperature for 6 hours, then gradually cooled down to lower temperature with different cooling rates for different tests. The growth was also stopped at different synthetic stages, with a series of photos taken *at once*, to help better understand the roles of individual specimens and the growth mechanism.

**2.4 Characterization of Black-P crystals:** X-ray diffraction of one selected crystal was carried out at room temperature on Rigaku SmartLab X-ray diffractometer equipped with CuK$_α$ radiation. Raman spectroscopy measurement was carried out on Thermo Scientific DXR Raman spectrometer. FEI Helios Nano Lab 400 was used to for SEM studies of the morphology and composition of the Black-P crystals and ternary clathrate compound. High resolution transmission electron microcopy (HRTEM, JEOL-2010F) images, selected area electron diffraction (SAED), and EDX mappings were executed on the sample of ultrasonically dispersed Black-P to characterize the structure and composition. Hall resistivity measurements were performed in a Quantum Design Physical Property Measurement System (PPMS) from 400 K down to 4 K through five probe configuration on a fresh sliced Black-P thin crystal.

### 3. Results and Discussion

Following synthetic route 2.2, the as-grown Black-P crystals in the quartz ampule tube are shown in the Fig. 1a. One can clearly see that the Black-P crystals form at the site where ternary clathrate $Sn_{24}P_{22-x}I_8$ originally resides, and no Black-P forms at the other end of the ampule (which has slightly lower temperature due to the small temperature gradient from center of the furnace to its edge), indicating the effective catalytic role of ternary clathrate $Sn_{24}P_{22-x}I_8$ for the Black-P growth. Both the optical images (Fig.1b) and SEM image (Fig. 1c) indicate that these crystalline Black-P are micro-ribbons morphology with uniform thickness and width. From the side view, the well-aligned layered-structure is clearly observed. And these crystals are easily-exfoliated into layered structure as shown in SEM images (Fig. 1d). The XRD pattern shows preferential orientation of *(0k0)* of the selected crystals, and the peaks are well indexed with no peaks from $Sn_{24}P_{19.3}I_8$ phase nor other impurities are observed. The Raman spectroscopy of the crystals show sharp peaks at 362.4, 440.2, and 466.8 cm$^{-1}$, which match well with the Raman shifts attributed to vibrations of the crystalline lattice of Black-P: $A_g^1$ (out of plane mode), $B_{2g}$ (in-plane mode), and $A_g^2$ (in-plane mode) phonon modes, respectively.[17]

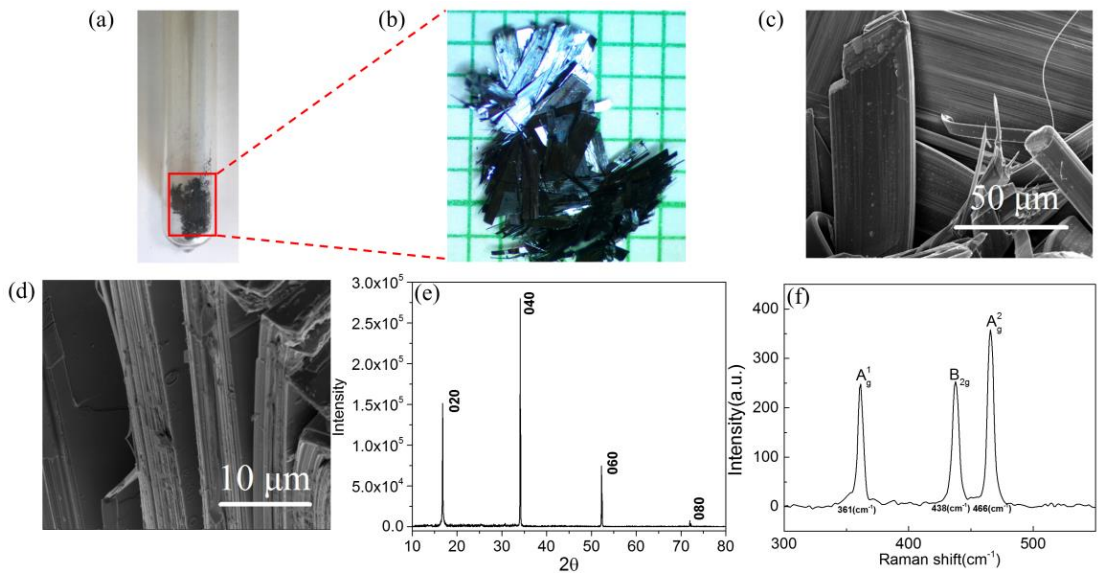

**Figure 1.** (a) Photos of Black-P crystals in the ampule tube grown directly from ternary $Sn_{24}P_{22-x}I_8$ compound; (b) enlarged view of Black-P crystals on the mm scale. (c) top view of SEM image of selected crystals showing ribbon-sharp morphology, (d) side view of SEM image showing the layered nature of the grown crystal; (e) XRD patterns, and (e) Raman spectrum of the grown Black-P crystals.

Our direct synthesis of Black-P through stable ternary Sn-P-I catalyst, rather than elemental components or gaseous $SnI_4$ and $SnI_2$ species, provide an alternative method for Black-P synthesis under relatively lower synthetic temperature and pressure. From the available literature[5,8-14] where the starting materials and temperature profile were precisely given, the target temperature we used here is slightly lower (550 °C comparing with 590 °C-650 °C), and more importantly, the pressure inside the ampule at the highest synthesis temperature is much lower. We have carried out several control experiments varying the amount of the starting materials, and found out the minimum Red-P can be as low as 80 mg, which is equivalent to ~ 0.5 MPa, by assuming sublimed $P_4$ vapor at the highest synthesis temperature following the ideal gas law (confined quartz ampule space is 10 mm inner diameter × 10 cm length). This pressure is much smaller than previously reported value of ~ 3.5 MPa (~500 mg Red-P sublimed at 650°C using a similar size of quartz ampule).[13] This will be beneficial for future large scale Black-P synthesis by other methods, e.g. chemical vapor depsotion, since the high pressure caused by limited ampule space (which could potentially break the quartz ampoule and cause safety issues if not handled properly) is now not a critical limiting factor any more.

Up to now, two phases of the ternary compounds consisting of Sn, P, and I are discovered.[15,16,18] One is $Sn_{24}P_{22-x}I_8$ with type-I clathrate structure[17,19] and the other is SnIP with double helices structure.[18] Recently, Nilges et. al. indicated that SnIP could decompose peritectically at 476 °C with formation of $Sn_{24}P_{22-x}I_8$ in an tightly enclosed space.[18] $Sn_{24}P_{22-x}I_8$ was used in our research for the growth of Black-P in this work. The $Sn_{24}P_{22-x}I_8$ belongs to a well-known type–I clathrate structure[17,19] with a 3D inversed network composed by Sn and P atoms as host framework, and I as guest atoms (Fig. 2a). The compound crystallizes in the cubic structure with space group $Pm\overline{3}n$(#223), with five different atomic Wyckoff positions. It has vacancies within the host framework, as stressed by the formulation $Sn_{24}P_{19.3}\square_{2.7}I_8$, where □ denotes a vacancy. Sn atoms at *24k* position, and P1 atoms at *16i* position, together forms two pentagonal dodecahedra. Another P2 atoms, forming the hexagonal faces of the 6 tetrakaidecahedrons, partially occupy the *6c* atomic positions (¼, 0, ½) which causes the vacancy structures in the framework (Fig. 2b and 2c). Two of the 8 iodide atoms are located at the center of the 2 pentagonal dodecahedrons (I1 in *2a* position) and the 6 others at the center of the 6 slightly larger tetrakaidecahedrons (I2 at *6d* position). Previous lattice dynamics analysis[16] has indicated that the host-guest interaction in this inverse-clathrate compound is weak and that the dynamics of the guest atoms can be treated independently of that of the host framework. Our controlled experiments and differential scanning calorimetry (DSC) measurements show that this phase decomposes irreversibly to gaseous $P_4$ (g), $SnI_2$ (g) and liquid Sn (l) above 620 °C in the closed ampoules. The ternary $Sn_{24}P_{22-x}I_8$ remains as solid at the highest synthesis temperature 550 °C as evidenced also in Fig. 2d. The unique structure vacancies at the P sites in the framework of $Sn_{24}P_{19.3}\square_{2.7}I_8$, might hold the glue for its catalytic effect to grow Black-P crystals through the direct interaction of phosphorus vapor with this solid ternary precursor.

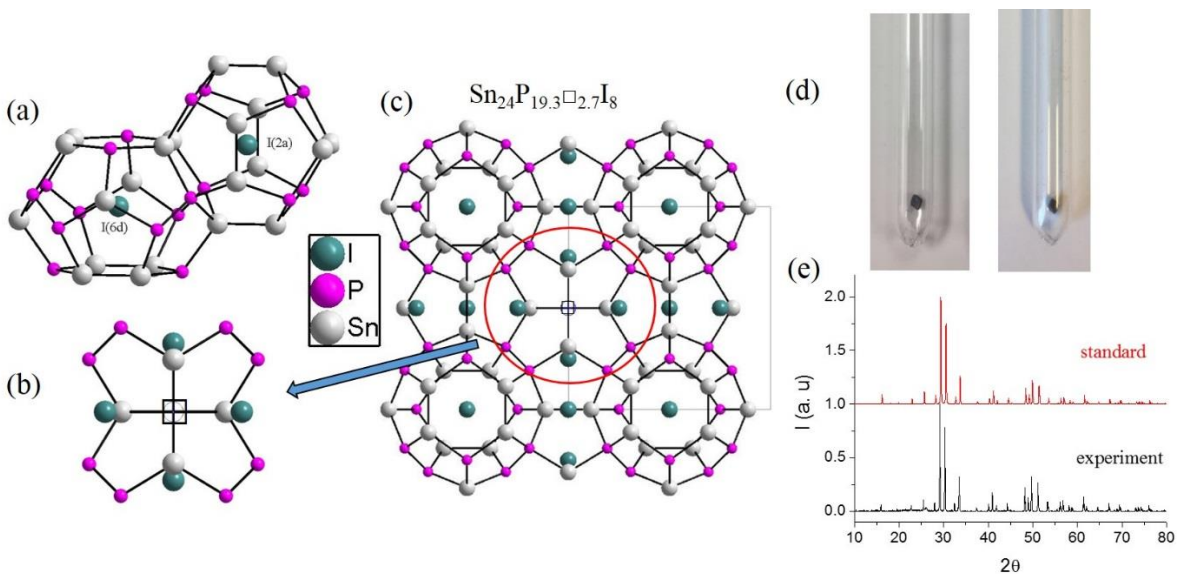

**Figure 2.** (a)Pentagonal dodecahedra and tetrakaidecahedron framework formed by Sn and P atoms in the clathrate $Sn_{24}P_{22-x}I_8$, where two different types of I atoms (2a and 6d) are located at the centers of each polyhedron. (b) and (c) The crystal structure of $Sn_{24}P_{22-x}I_8$, where the vacancy □ caused by partially occupied P2 atom is highlighted. (d) Pictures of $Sn_{24}P_{22-x}I_8$ pellet inside reaction ampule before and after 550 °C, showing $Sn_{24}P_{22-x}I_8$ remains unchanged. (e) X-ray powder pattern of $Sn_{24}P_{22-x}I_8$ used for section 2.2.

To investigate thoroughly the growth mechanism of Black-P crystals, we also grew Black-P crystals through the elemental Red-P, Sn, $I_2$ and binary Sn-I compounds. We have varied Sn:I ratio and the amount of Red-P used for the synthesis to investigate the necessity of chemicals and pressure required for the growth of Black-P. In addition, we also compare the synthetic attempts of using tube furnace with hot/cold ends (temperature gradient > 100 °C which promote vapor transport reactions) and box furnace with uniform temperature (temperature gradient < 5-10 °C from the center of furnaces to the edge of front door) which will be helpful for us to understand the possible growth mechanism as well. The growth processes were also stopped at different synthetic stages, with a series of photos taken *at once*, and products analyzed afterwards. Some crucial results are listed in the Table 1 and 2, from which we have found out: 1) solely chemical additive of $SnI_2$ or $SnI_4$ does not work. Excess Sn besides the formation of $SnI_2$ is a crucial ingredient for a successful growth. Sn:$I_2$ > 2:1  atomic ratio was found out to work the best (Table 1, Recipes 1-5); 2) Black-P crystals could be grown from both tube furnace and box furnace from elements. Temperature gradient (> 50 °C) in tube furnaces are found out be helpful to facilitate Black-P crystals growth (if grown from elements) but not a necessity; 3) Black-P crystals form during cooling process; $Sn_{24}P_{22-x}I_8$ forms first between 600 - 520 °C, then Black-P forms after temperature decreases below 520 °C; 4) Black-P starts to form when temperature is below 500 °C (500 °C and 480 °C) with evidence that small segment of Black-P emerges at the end of the tube. However, at this stage, most of the P remains as $P_4$ (g), and is transferred to Red-P (coated inside the tube) and White-P droplets on the quartz wall if reaction is quenched; 5) The formation of Black-P is nearly completed at the temperature below 420 °C with evidence as following: a) the quartz tube is clean clear, indicating nearly no $P_4$ (g) transfer to Red-P or White-P upon quenching; b) the crystals' shapes and morphology are nearly identical for the two batches quenched at 420 °C and 350 °C (meaning formation is completed at the T ~ 420 °C). Therefore, it is fair to say that the formation of Black-P is a rapid growth progress between 420 and 520 °C.

*Table 1: Control experiments for formation of Black-P using a slow cooling rate*

| No. | Reactants | Max. T (°C) | Black-P formation |
|---|---|---|---|
| 1 | $RP+SnI_4$ | 600 | no |
| 2 | $RP+SnI_2$ | 600 | no |
| 3 | $RP+Sn+I_2$ (Sn: $I_2 \leq 1:1$) | 600 | no |
| 4 | $RP+Sn+SnI_2$ | 600 | yes |
| 5 | $RP+Sn+I_2$ (Sn: $I_2 > 1:1$) | 600 | yes |
| 6 | $RP+Sn_{24}P_{19}I_8$ | 500-600 | yes |
| 7 | $RP+ Sn_{24}P_{19}I_8$ | 480 | no |

*Table 2: Control experiments for formation of Black-P crystal at different quenching temperature after slowing cooling from 600 °C. RP: Red-P, WP: White-P, and BP: Black-P.*

| Quenching temperature | 600°C | 550°C | 520°C | 500°C | 480°C | 420°C | 350°C |
|---|---|---|---|---|---|---|---|
| Black-P formation | No | no | no | yes | yes | yes | yes |
| Products | RP+WP | Sn-P-I +RP+WP | Sn-P-I + RP+WP | BP+ RP | BP+RP | BP | BP |

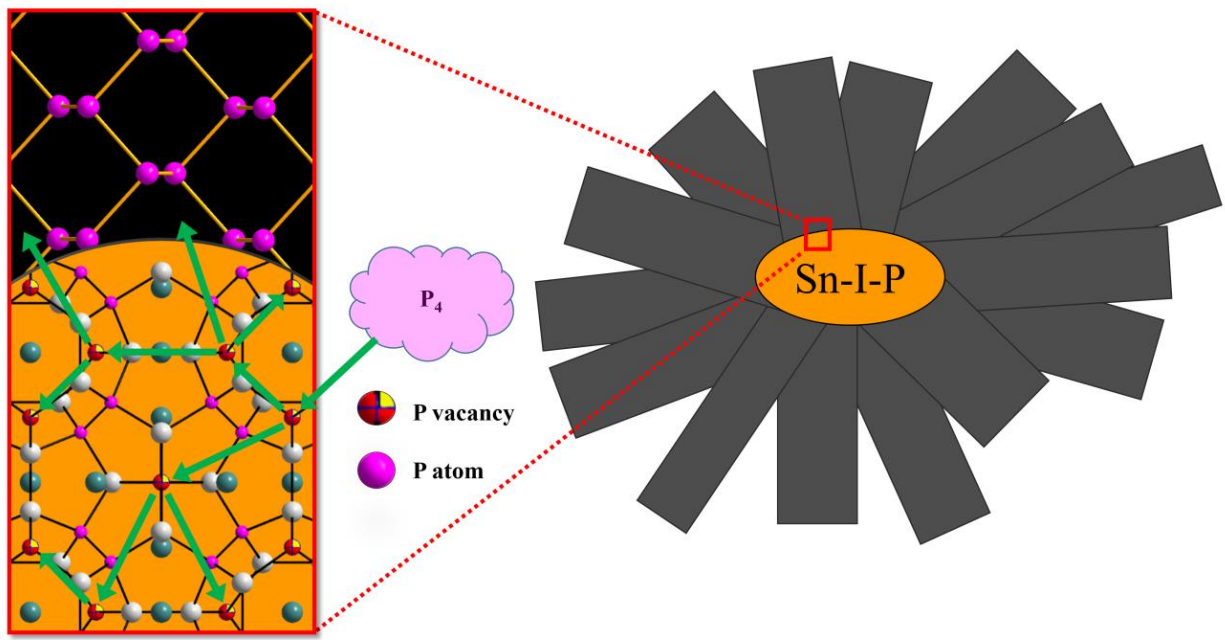

**Figure 3**. Proposed growth mechanism. The orange part represents ternary compound of $Sn_{24}P_{22-x}I_8$; the black part represents Black-P; the purple dots represent P atoms; the colorful dots represent P vacancies; the green dots represents I atoms and grey dots represents Sn atoms. The green arrows indicate the possible diffusion path of P atoms with the assistance of P vacancies.

Based on the aforementioned findings, we could conclude with a likely Black-P crystal growth mechanism where the role of individual element at different temperature stages are identified. In the growth of Black-P with elements or Sn-I binary compounds, the initial materials of Sn, $I_2$ and Red-P are readily reacted/vaporized to form gaseous $SnI_2$, and $P_4$ upon heating from room temperature to 600 °C. The excess of Sn remains as liquid at this stage, but could be transported to the cold side (if any) through vapor transport reactions ($Sn + I_2 \longrightarrow SnI_2$). Upon cooling, from 600 °C to ~ 500 °C, these gaseous species react with each other and form the solid ternary $Sn_{24}P_{22-x}I_8$ phase first, and the excess of P element still remains as the $P_4$ vapor. During this stage, the chemical vapor transport reaction does help with the formation of solid phase of $Sn_{24}P_{22-x}I_8$, but no Black-P crystal has formed yet. Upon further cooling below ~ 520 °C, the $P_4$ vapor steadily transfers to Black-P crystals through a likely vapor-solid-solid (VSS) mechanism[20-22] via $Sn_{24}P_xI_8$ phase. During this stage, the $Sn_{24}P_{22-x}I_8$ phase not only provides the nucleation center for Black-P crystals but also acts as a media to transport elemental P for the continuous Black-P crystal growth. The host-guest and P-site vacancy structure features in the clathrate $Sn_{24}P_{22-x}I_8$ phase allow chemical reactions of $P_4$ vapor with the host structure where the excess P then could diffuse out from the host structure (therefore suppress the energy barrier) to form Black-P crystals. The whole reaction happens rather fast (less than 12 hours) and stopped at the temperature below 350 °C.

It is worthwhile to note that vapor-phase mechanisms including vapor-liquid-solid (VLS) and Vapor-Solid-Solid (VSS) are primary growth modes for nanowires and nanotubes,[23-28] where vapor is the precursor of the final product (solid), and catalyst is a nano droplet (liquid) or nanoparticle (solid). For VSS mode, the solid catalyst is either an alloy or a compound consisting foreign and final product elements. In the growth of Black-P by $Sn_{24}P_{22-x}I_8$ catalyst, VSS is the possible growth mechanism with the evidences: 1) all as-grown Black-P belts attach on $Sn_{24}P_{22-x}I_8$ particles; 2) in the growth temperature of Black-P, $Sn_{24}P_{22-x}I_8$ is a solid; 3) in the lattice of $Sn_{24}P_{22-x}I_8$, P vacancies can vary from X1% to X2%, depending on the partial pressure of $P_4$. The vacancies in a solid may greatly enhance the diffusion of the certain atoms in it.[29] Therefore, we propose that our Black-P growth is governed by the VSS mode, as shown in Fig. 3. When temperature is lower than 520 °C, $Sn_{24}P_{22-x}I_8$ is stable, then $P_4$ can decompose at the surface of $Sn_{24}P_{22-x}I_8$ and P atoms diffuse into $Sn_{24}P_xI_8$. When the P concentration is high enough in $Sn_{24}P_{22-x}I_8$, P begins to segregate at some locations on the surface of $Sn_{24}P_{22-x}I_8$ as Black-P, the most stable phase for phosphorus in thermodynamics. With time going on, the Black-P P grows longer. Different from VLS, in general, the nanowires grown under VSS have uniform diameter from tip to bottom, since the size of catalyst in solid is more stable. In this work, the uniform width of Black-P belt is also uniform, which supports VSS hypothesis.

Finally, we examined the temperature dependence of the carrier mobility through Hall measurements. Six wire method was utilized to simultaneously measure the resistivity and Hall coefficient of freshly cleaved bulk Black-P crystals. The inset of Fig. 4 schematically shows the contact of 6 wires, where contact leads of 1, 2, 5, 6 were used to determine the electric resistivity of the sample and contact leads of 1,3,4,6 were used for the Hall measurement. The contribution of the transvers resistance $R_{xx}$ which was induced due to any possible misalignment of the two Hall probes, is further eliminated through $R_{xy}(H)=(R_{xy}(+H)-R_{xy}(-H))/2$, since $R_{xx}$ will not change when the direction of applied magnetic field is changed to the opposite direction. Through this method, we can get the actual Hall resistance of the bulk sample. The positive Hall coefficient indicate the charge carriers are mainly p-type in our bulk Black-P samples. The derived bulk Hall mobility shows a broad peak around 250 K, with the room temperature mobility μ is near 350 $cm^2/V \cdot s$ as shown in Fig. 4. The obtained mobility value is higher than most of the reported value of bulk crystals,[30] and is even comparable with some of reported value with few layers and monolayer of Black-P.[31-33] Given the unavoidable micro-cracks on the surface during cleavage (which could be clearly observed in the figure inset), we believe that the actual value of mobility could be even higher, which further support the high quality of the Black-P crystals synthesized through this new method.

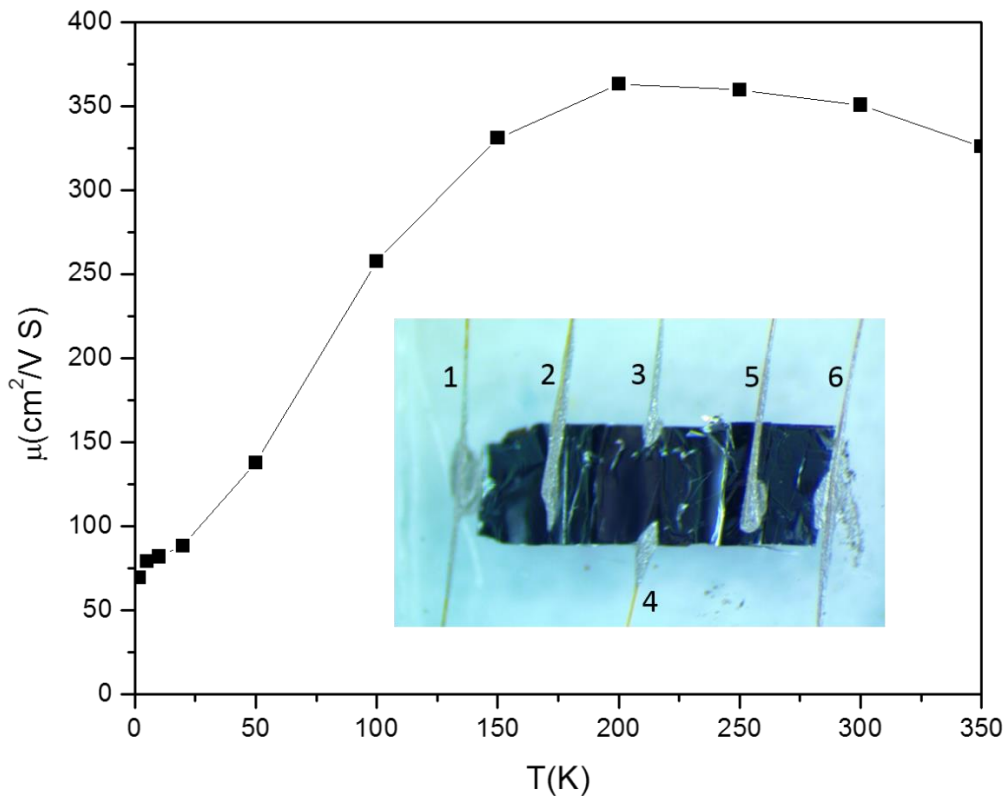

Figure 4. Hall mobility of a Black-P belt with temperature along a randomly chosen direction

In conclusion, we have synthesized high quality Black-P crystals *in-situ* using ternary clathrate $Sn_{24}P_xI_8$ as the catalyst. X-ray diffraction, SEM, Raman and TEM studies show high quality of grown crystals. Through detail investigation of growth mechanism using different synthesis routes, we have found out the growth temperature and pressure range for Black-P crystal growth, identified the role of individual chemical additives in favor of the formation of Black-P, and suggested that the VSS, rather than the commonly accepted vapor transport reaction, is a likely growth model. The P vacancy in the framework of ternary clathrate is apparently crucial for the *in-situ* growth of Black-P at clathrate site. The Hall measurement shows high mobility in the as-grown bulk crystals.

**Acknowledgments:** This work in UT Dallas is supported in part by US Air Force Office of Scientific Research Grant No. FA9550-15-1-0236, and the Start-Up funds from University of Texas at Dallas. Q. Yu thanks the support from 2D Carbon Technology, and the Start-up funds from Texas State University and X. Fan thanks the support from China Scholarship Council.